\documentclass[a4paper]{mem}
\usepackage{graphicx}
\begin{document}
   \title{Near-Infrared photometry of LMC cluster Reticulum
}

   \author{M. Dall'Ora \inst{1,2}, 
  G. Bono \inst{2}, J. Storm \inst{3}, V. Testa \inst{2},
  G. Andreuzzi \inst{2}, R. Buonanno \inst{1,2}, F. Caputo \inst{2},
   V. Castellani \inst{2,4}, C.E. Corsi \inst{2}
   S. Degl'Innocenti \inst{5}, G. Marconi \inst{2,6},
   M. Marconi \inst{7}, M. Monelli \inst{2} \and V. Ripepi \inst{7}\fnmsep
}

   \offprints{M. Dall'Ora,
   \email{dallora@mporzio.astro.it}}
\mail{via Frascati 33, 00040 Monte Porzio Catone }

   \institute{Dipartimento di Fisica, Universit\`a di Roma Tor Vergata,
     Via della Ricerca Scientifica, 1, 00133 Rome, Italy 
   \and  INAF, Osservatorio Astronomico di Roma, via Frascati 33 Monteporzio
      Catone, Rome, Italy
   \and Astrophysikalisches Institut Potsdam, An der Sternwarte 16,
      D-14482 Potsdam, Germany
   \and INFN, Sezione di Pisa, via Livornese 582/A, S. Piero a Grado, 56010, 
     Pisa, Italy
   \and Dipartimento di Fisica, Universit\`a di Pisa,
     piazza Torricelli 2, 56126 Pisa, Italy
   \and ESO, Alonso de Cordova 3107, Vitacura, Santiago , Chile
   \and INAF, Osservatorio Astronomico di Capodimonte, via Moiariello 16, 
     80131 Napoli, Italy
      }

\noindent

   \abstract{We present near-infrared ($JK_s$) time series data of the Large
      Magellanic Cloud (LMC) cluster Reticulum. The observing strategy and 
      data reduction (DAOPHOTII/ALLFRAME) allowed us to reach a photometry 
      accuracy of the order of 0.02 mag at limiting magnitudes typical 
      of RR Lyrae stars. We are interested in Reticulum, since it hosts a 
      sizable sample of RR Lyrae (32), and therefore the use of the $K$-band
      Period-Luminosity-Metallicity ($PLZ_K$) relation will allow us to supply
      an accurate LMC distance evaluation. The main advantages in using this 
      method is that it is marginally affected by off-ZAHB evolutionary 
      effects and by reddening corrections. As a preliminary but robust 
      result we find a true distance in good agreement with the 
      LMC Cepheid distance scale, i.e. $\mu = 18.47 \pm 0.07$ mag.
      
   }
   \authorrunning{M. Dall'Ora et al.}
   \titlerunning{Near-Infrared Photometry of LMC cluster Reticulum}
   \maketitle

\section{Introduction}
RR Lyrae stars are traditionally considered good Population II distance 
indicators, since they approximatively share a common $V$ magnitude,
are easily detectable from their light curve and they are bright enough to 
be observable at large distances. The mean $V$ magnitude of RR Lyrae does 
depend on the metallicity, i.e. $M_V = \alpha [Fe/H] + \beta$. Unfortunately, 
there is no general consensus on the slope of this relation, and current 
values range from $\alpha \approx$ 0.18 to $\approx$ 0.30. When applied to 
globular clusters, these distance uncertainties imply a substantial 
uncertainty on the absolute age estimate as well as on the statement of a 
quantitative age-metallicity relation. Finally, we note that it has 
been suggested that the $M_V$ vs. [Fe/H] relation might not be strictly linear
when moving from metal-poor to metal-rich RR Lyrae (Caputo et al. 2000).
Moreover, even in the same globular cluster (i.e. in a population of stars 
characterized by the same metallicity), evolutionary effects
can produce a spread in magnitudes ranging from $\approx 0.2$ to $\approx
0.5$ mag (Sandage, 1990). 

These problems can be overcome in the near-infrared bands, because RR Lyrae 
in the $K$-band (2.2 $\mu m$) obey a well-defined period-luminosity relation
(Longmore et al. (1986, 1990). The
major advantages of such a relation are that $K$-band magnitudes are only
marginally affected by both metallicity and reddening effects. 
In their theoretical work, Bono et al. (2001) derived a $PLZ_K$ relation 
for the $K$-band. They found that the relation shows only a mild dependence on 
the luminosity level and a small dependence on the mass. 
Moreover, Bono et al. (2003) showed that the residual uncertainty on the 
luminosity level can be further reduced by  using a period-color-metallicity 
relation. The theory has been tested by Bono et al. (2002) on RR Lyrae itself, 
whose trigonometric parallax was recently measured by HST 
(Benedict et al., 2002). Bono et al. found a ``pulsational parallax" of 
$3.858 \pm 0.131$ mas, to compare with $3.82 \pm 0.20$ mas obtained by Benedict 
and coworkers.

In the present work we apply these theoretical relationship to RR Lyrae stars
hosted in the LMC cluster Reticulum, since LMC is the cornerstone of the 
extragalactic distance scale.

%
   \begin{figure*}
   \centering
   \resizebox{\hsize}{!}
      {\includegraphics[clip=true]{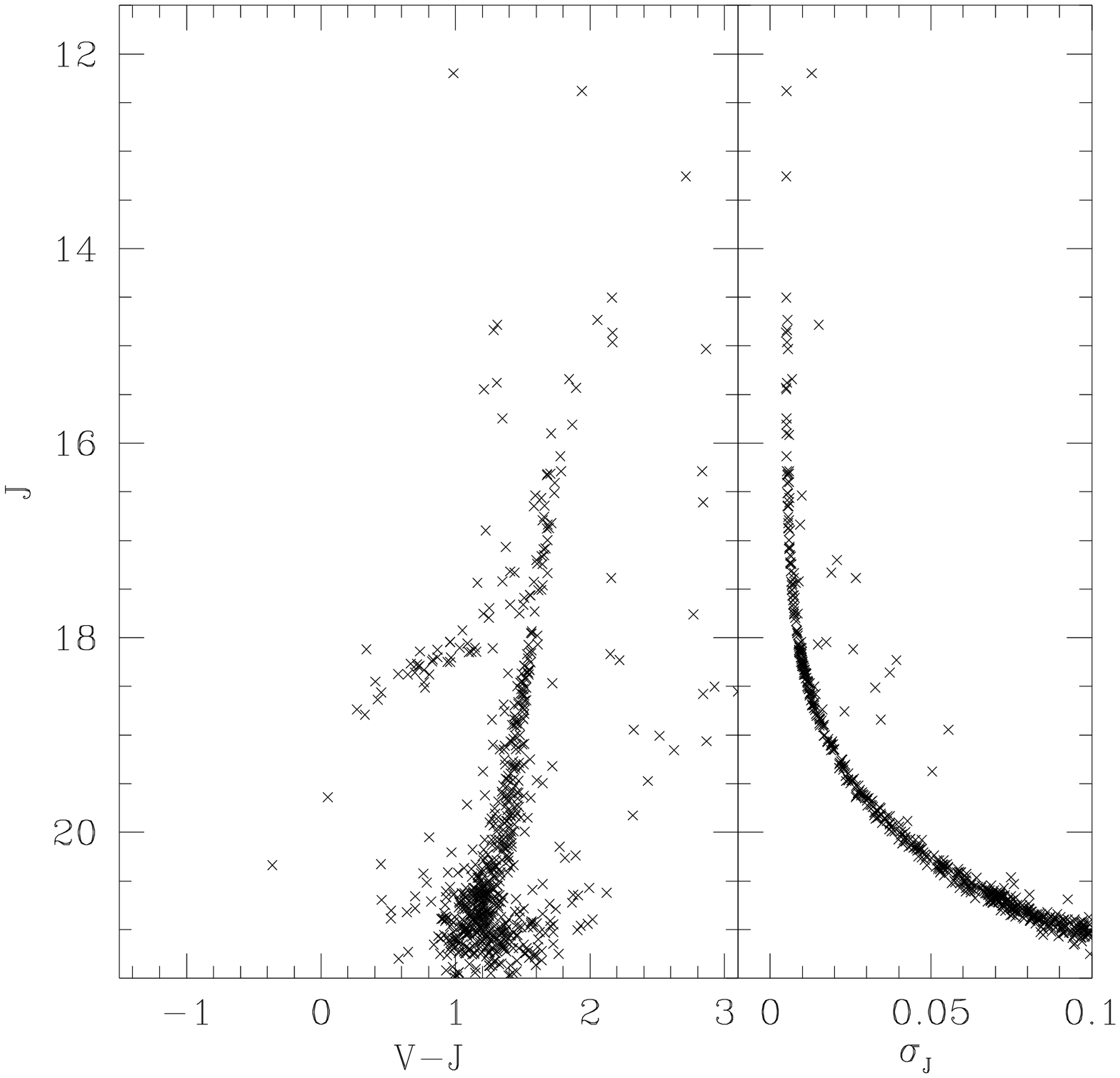}
   \includegraphics[clip=true]{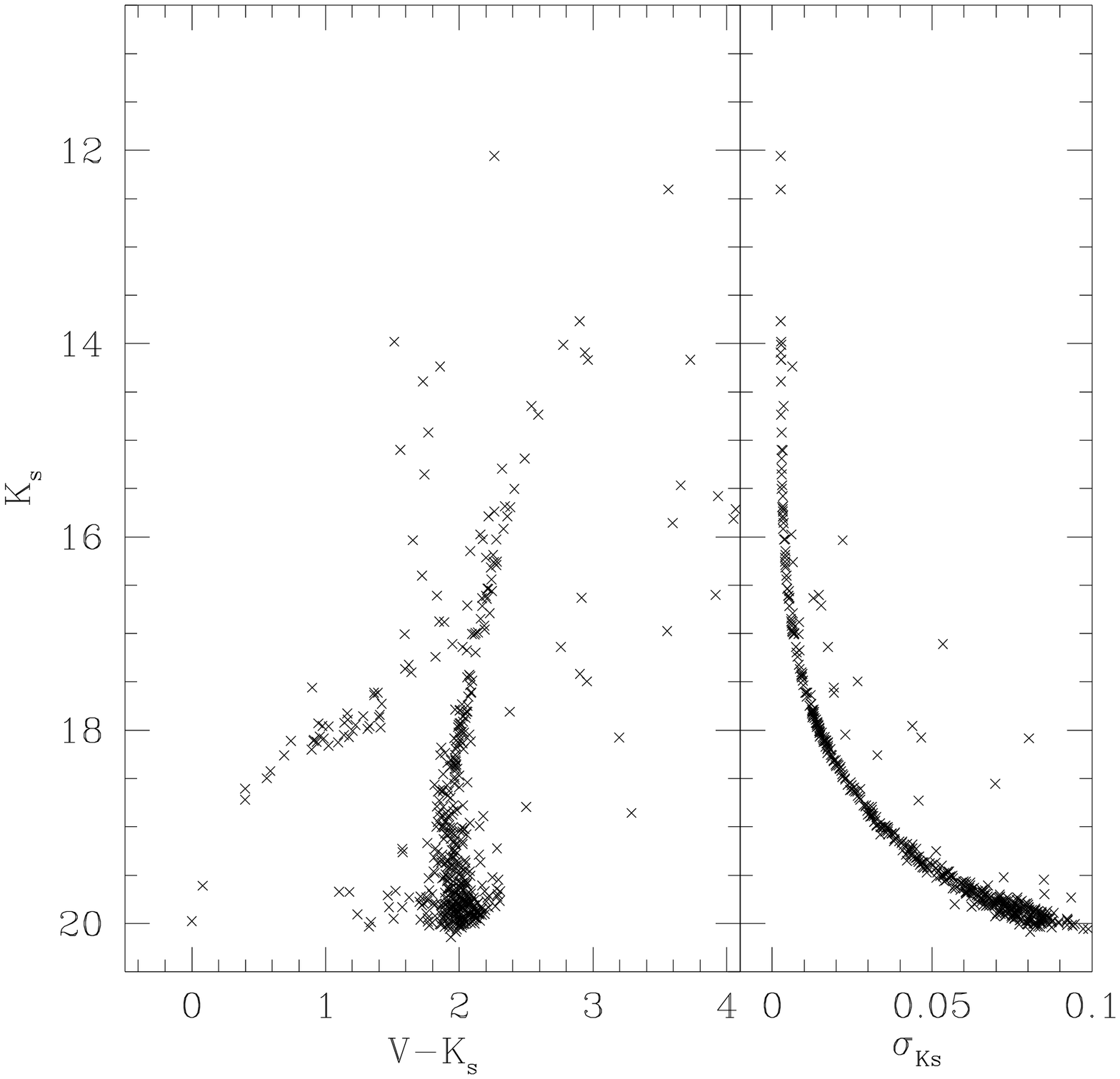}}
     \caption{$J-(V-J)$ (left) and $K_s-(V-K_s)$ color-magnitude diagrams of
     Reticulum. ALLFRAME mean magnitudes are plotted. We detected $\approx 900$
     stars. The two figures only show stars with DAOPHOT parameters
     \textit{sharpness (sh)} and $\chi$ (that are diagnostics of the goodness
     of the photometry) in the range $-1 \le sh \le 1$ and $\chi < .6$, 
     respectively.}  
        \label{fit1}
    \end{figure*}
%

\section {Data analysis}
$J$ and $K_s$ images were collected in three different runs with SOFI@NTT/ESO 
from December 1999 to February 2002, and they were pre-reduced, following the
procedures outlined in the SOFI user's manual (Lidman et al., 2002), by means of 
standard IRAF commands. Photometric reduction on 46 $J$ and 171 $K_s$ useful 
frames was performed with DAOPHOTII/ALLFRAME packages (Stetson, 1987, 1994). 
Photometric zero-point was estimated on the Persson et al. (1998) 9109 
standard star.

$J -(V-J)$ and $K_s -(V-K_s)$ color-magnitude diagrams were obtained by
coupling infrared data with $V$ observations collected with SUSI1 and 
SUSI2@NTT/ESO (Fig.1). Data plotted in Fig. 1 are ALLFRAME mean magnitudes.
Note that the photometric accuracy for RR Lyrae stars is better than 0.02
mag, and the good photometric accuracy is also supported by the narrow
distribution of stars along the Red Giant Branch. Note that we reached limiting 
magnitudes of $\approx 21$ and $\approx 19.7$ mag in $J$ and $K_s$ bands, 
respectively.

\section{Distance modulus}
To accomplish our goal we only used RR Lyrae stars with well-sampled 
light curves, namely 21 stars in the J and 23 in the $K_s$ band. First 
overtone  pulsators were ``fundamentalized", i.e. we added 0.127 to their 
$\log P$, to use the same $PLZ_K$ relation for fundamental and first overtone 
pulsators. Fig.2 displays observed $J$ (left panel) and $K_s$ magnitudes vs. 
$\log P$, showing the well-defined correlation between these two quantities.
In the following we will also put $K = K_s$, since the color term in the 
transformation between these two filters is negligible (Lidman et al., 2002).

For the $K$ band we used the fundamental pulsators relation given by 
Bono et al.  (2003)

\begin{eqnarray}
M_K = 0.568 - &2.101 \log P + 0.125 \log Z{}\nonumber \\
& - 0.734 \log L/L_\odot
\end{eqnarray}

The luminosity of each pulsator can be evaluated using the $(V-K)$ 
period-color relation (Bono et al., 2003)

\begin{eqnarray}
(M_V - M_K) = 3.963 + 1.986 \log P + {} \nonumber \\
  +0.162 \log Z -1.662 \log L/L_\odot
\end{eqnarray}

Adopting $\log Z=-3.4$ (Suntzeff et al., 1992), $E(B-V)=0.02$
(Walker, 1992) and $A_K=0.11 A_V$ (Cardelli et al., 1989) we obtained a  
true distance modulus

\begin{eqnarray}
&(DM)_{K,0}=18.55 \pm 0.07 \nonumber
\end{eqnarray}

We also estimated the distance with a period-luminosity-metallicity relation
for the $J$-band

\begin{eqnarray}
M_J = 1.669 - &1.491 \log P + 0.048 \log Z- {} \nonumber \\
&-1.251 \log L/L_\odot
\end{eqnarray}
\noindent
where once again we estimated the luminosity with equation (2).
Adopting $A_J=0.28 A_V$ (Cardelli et al., 1989) we obtained

\begin{eqnarray}
&(DM)_{J,0}=18.51 \pm 0.06 \nonumber
\end{eqnarray}

Current distance estimates agree quite well, within the errors, with distance 
determinations available in the literature and in particular with the 
distance evaluation based on LMC Cepheids, i.e. ($\mu=18.55 \pm 0.06$, Fouqu\`e 
et al., 2003). Note that, if we assume a difference in distance
between Reticulum and the LMC center of $\Delta \mu = 0.08$ (Walker, 1992), 
then the true distance to LMC becomes $(DM)_{J,0}=18.43 \pm 0.06$ and 
$(DM)_{K,0}=18.47 \pm 0.07$, respectively. Both values are in good
agreement with the distance modulus of LMC evaluated by Cacciari \& Clementini
(2003), namely $\mu=18.48 \pm 0.05$, obtained by averaging different absolute 
magnitude determinations for RR Lyrae stars.

%
   \begin{figure*}
   \centering
   \resizebox{\hsize}{!}
   {\includegraphics[clip=true]{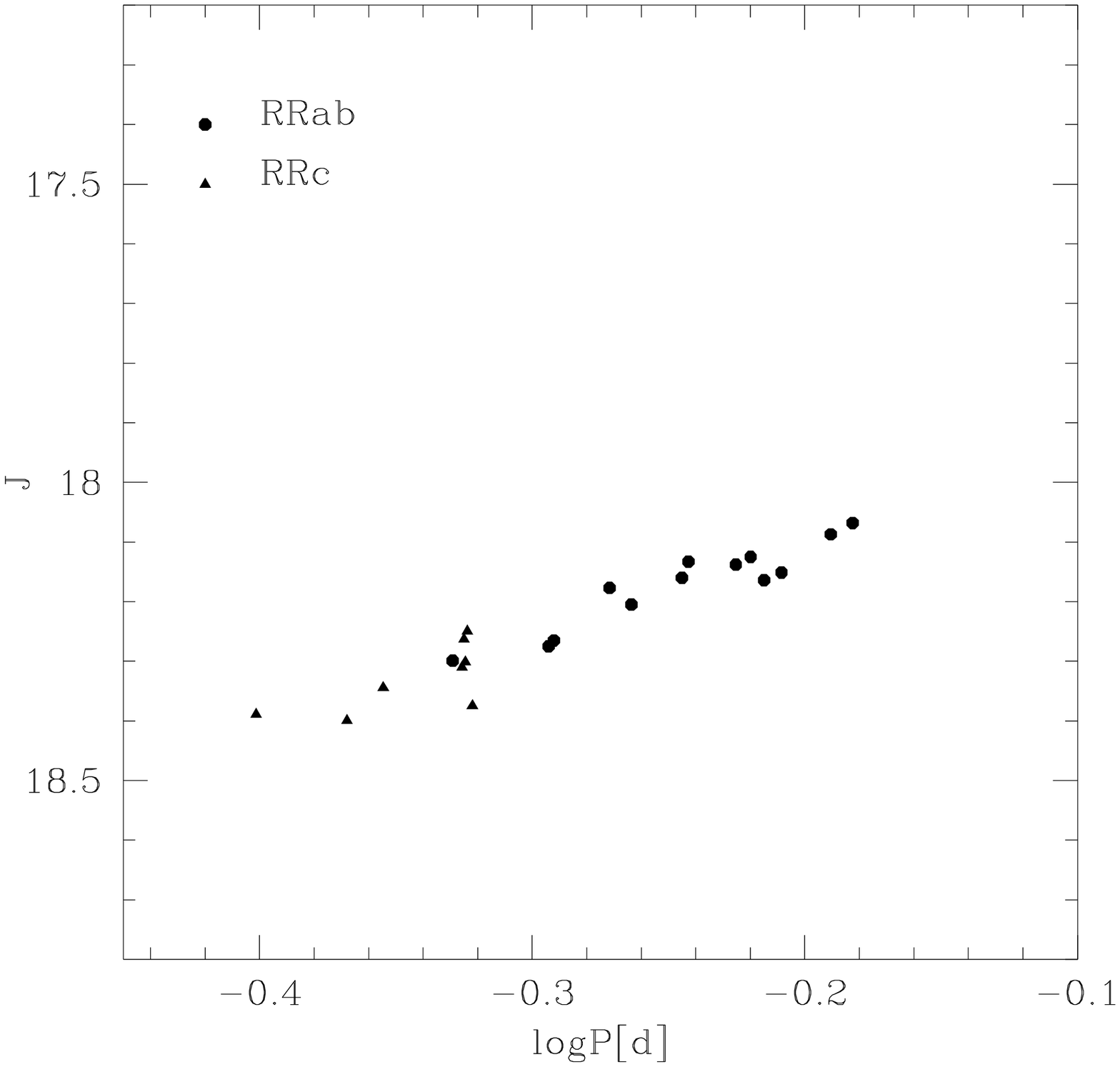}
   \includegraphics[clip=true]{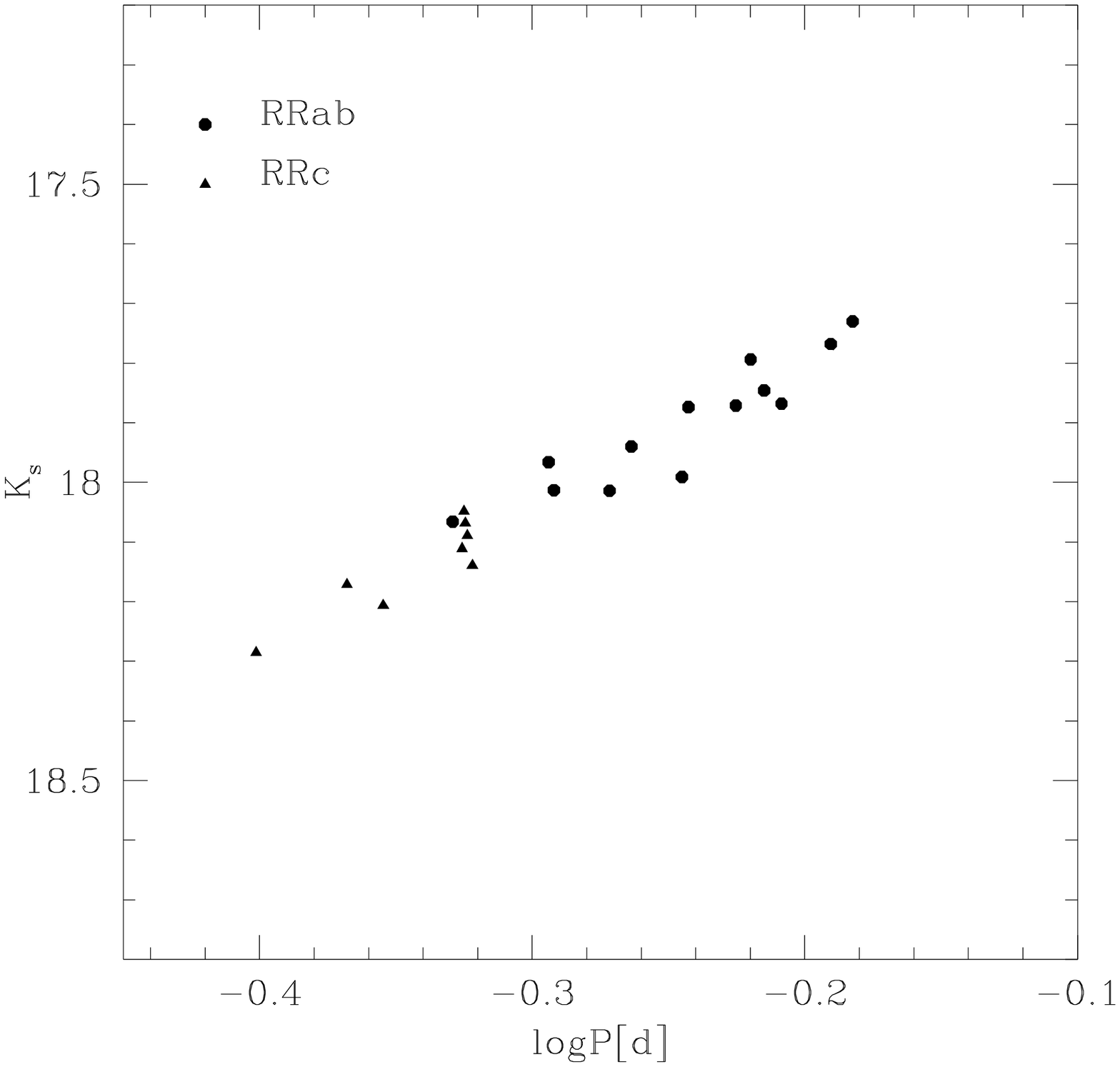}}
   \caption{Observed period-luminosity relations for $J$ (left panel) and
   $K_s$-bands. Filled circles depict fundamental pulsators (RRab), while 
   filled triangles are first overtone pulsators (RRc). Periods of RRc have
   been ``fundamentalized" (see text).}
\end{figure*}
%

\section{Conclusions}
Empirical and theoretical evidence suggest that the $PLZ_K$ relation is 
a quite promising method to obtain accurate distance estimates for RR Lyrae 
stars. With this
relation we obtained a distance modulus to the LMC cluster Reticulum that is
in good agreement with the LMC Cepheid distance. The accuracy of 
the result could be further improved by fitting the light curves with suitable
templates (Jones et al., 1996). Finally, as a preliminary result, we also 
estimated the Reticulum distance modulus with a period-luminosity-metallicity 
relation for the $J$-band, and we obtained a value in good agreement with 
the $PLZ_K$ distance.

%

\bibliographystyle{aa}

\end{document}